\documentclass{Interspeech}

\interspeechcameraready 

\usepackage{graphicx}
\usepackage{amsmath}
\usepackage{booktabs}
\usepackage{subcaption}

\title{Towards Early Prediction of Self-Supervised Speech Model Performance}

\author[affiliation={1}]{Ryan}{Whetten}
\author[affiliation={1,2}]{Lucas}{Maison}
\author[affiliation={3}]{Titouan}{Parcollet}
\author[affiliation={4}]{Marco}{Dinarelli}
\author[affiliation={1}]{Yannick}{Estève}

\affiliation{Laboratoire Informatique d’Avignon}{Avignon Université}{France}
\affiliation{}{Thales SIX}{France}
\affiliation{}{Samsung AI Center Cambridge}{United Kingdom}
\affiliation{Laboratoire d'Informatique de Grenoble}{Université Grenoble Alpes}{France}
\email{ryan.whetten@univ-avignon.fr}

\usepackage{comment}

\begin{document}

\maketitle

\begin{abstract} 
In Self-Supervised Learning (SSL), pre-training and evaluation are resource intensive. In the speech domain, current indicators of the quality of SSL models during pre-training, such as the loss, do not correlate well with downstream performance. Consequently, it is often difficult to gauge the final downstream performance in a cost efficient manner during pre-training. In this work, we propose unsupervised efficient methods that give insights into the pre-training quality of SSL speech models, namely, measuring the cluster quality and rank of the embeddings produced by the SSL model. Results show that measures of cluster quality and rank correlate better with downstream performance than the pre-training loss, reducing the need for GPU hours and labeled data in SSL model evaluation.
\end{abstract}

\section{Introduction}
\label{sec:intro}
Self-Supervised Learning (SSL) is the process of training a model where the targets are derived directly from the input data itself, decreasing the need of human-labeled data.
SSL models are usually trained in two stages: \emph{pre-training}, where the model is trained using unlabelled data, and subsequently, \emph{fine-tuning}, where the model is trained for a specific task using labeled data~\cite{mohamed2022self}. In the speech processing domain, SSL models have performed well on a variety of tasks such as Automatic Speech Recognition~(ASR) or Speaker Verification (SV)~\cite{mohamed2022self,yang2021superb}.

Although performant, SSL speech models are computationally and memory expensive, usually requiring anywhere from 8~-~128 high-end GPUs~\cite{chen2023reducing, babu2021xls}, and up to tens of millions of hours of data~\cite{zhang2023google}. While recent work has been done to reduce the computation and memory requirements necessary to train these models~\cite{chen2023reducing, baevski2023efficient, parcollet2023efficiency, whetten2024analysis}, limited research has focused on the efficient evaluation of these models during pre-training. Nevertheless, if it were possible to evaluate a model's performance effectively during pre-training, the total number of models requiring full training could be significantly reduced, leading to significant energy and compute savings.

While it is common practice to monitor the SSL pre-training loss, it does not necessarily correlate well with the downstream performance~\cite{chen2023reducing}. Thus, the most reliable method of evaluation remains to simply test each model on the downstream tasks. This can be done by fine-tuning for a specific task, or by using benchmarks such as SUPERB or the Multi-Probe Speech Self-Supervision benchmark (MP3S)~\cite{yang2021superb, zaiem23b_interspeech}, in which the pre-trained SSL weights are frozen, and only a light downstream model, sometimes called a \emph{probe}, is trained. Though these evaluation methods are feasible, they require labeled data and additional GPU hours.

To overcome the need for labeled data and additional GPU hours, researchers in the field of computer vision proposed using the effective rank to evaluate the quality of SSL representations~\cite{rankme, entropicRank}. This methodology has shown promising results in the speech domain on the tasks of Phoneme Recognition, Keyword Spotting, and Speaker Identification~\cite{aldeneh2024towards}. 
However, whether this methodology extends to other tasks such as ASR and SV remains an open question. Additionally, no work has explored other measures like clustering methods. Furthermore, whether this approach can be used early in pre-training to predict final downstream performance remains unclear.

To address these gaps, we propose two compute- and data-efficient unsupervised methods as indicators of final downstream performance at early stages of pre-training: measuring the clusters quality and the rank of the embeddings outputted by SSL models. To test these methods, we experiment with the BERT-based Speech pre-Training with Random-projection Quantizer (BEST-RQ) architecture~\cite{pmlr-v162-chiu22a}, training 30 BEST-RQ models varying the batch size and the proportion of audio masked and focus on ASR and SV tasks.\footnote{code at https://github.com/whettenr/ssl-speech-evaluation} Results show that, for ASR, clustering measures and the rank of SSL embeddings correlate better with the final performance than the pre-training loss, even at early stages in training, having the potential to save thousands of GPU hours in future experiments.

\section{Background}
\label{sec:background}
In this section, we give an overview of BEST-RQ, explain how the pre-training loss is calculated, and show theoretically why it is not a good indicator of downstream performance. This is supported empirically in Section~\ref{sec:results}.


\subsection{Architecture of BEST-RQ}
\label{sec:sss-arch}
We use BEST-RQ due to its efficiency, performance, and open source implementation in the SpeechBrain library~\cite{whetten2024open, speechbrain}. In BEST-RQ there are three main components: the quantizer, the acoustic feature extractor (AFE) consisting of two CNN layers, and the context encoder made up of a series of Conformer layers~\cite{gulati2020conformer}. 
Given an audio, Mel filterbanks features are calculated and then given to the quantizer. The AFE also receives the Mel filterbanks features, except a random portion of Mel filterbanks are \emph{masked} (replaced with random noise). The output of the AFE is then passed to the context encoder.

Using a random frozen linear projection and a codebook lookup, the quantizer generates discrete pseudo-targets and the context encoder is trained to predict these targets for the masked portions. The loss is calculated by taking the cross entropy between the predicted targets and the pseudo-targets. Due to these training conditions, the context encoder must use the surrounding unmasked context in order to make accurate predictions on the masked portions~\cite{pmlr-v162-chiu22a}.

\subsection{Pre-training loss vs downstream performance}
\label{sec:loss-vs-ds}


In this section, we show theoretically how certain hyper-parameters changes can affect the loss. Let us consider two SSL BEST-RQ models $M_a$ and $M_b$, with respective masking hyper-parameters $m_a$ and $m_b$, which represent the percentage of audio that is masked during pre-training. Let $m_a < m_b$ and $m_a \approx 0$ meaning almost none of the input is masked. Let both models use the same quantizer $Q$. Finally, let $\text{x} = [x_1, x_2, ... x_T]$ be an input sequence of vectors of length $T$.

As the context encoder uses the global context to predict the targets of the masked sections, the larger the mask, the smaller amount of uncorrupted input available to the model.
Thus, the model $M_b$, having a larger mask $m_b$, will have more uncertainty in its prediction, resulting in higher entropy, and consequently, a higher loss. However, as will be explained subsequently, having a higher loss is not always a negative factor in this scenario.


Since $m_a$ is very small, $M_a$ would be trained to mimic $Q$, essentially learning a single linear projection—a simple task for Conformers—resulting in a low loss. In other words, as $M_a$ is trained, the output of $M_a(x_i) \approx Q(x_i)$. Conversely, because $M_b$ has a higher mask percentage, it does more than simply replicate $Q$. Instead, it uses the unmasked context to make accurate predictions of $Q$, compelling $M_b$ to learn the statistical properties of speech and audio. As a result, while $M_b$ incurs a higher loss than $M_a$, it learns important statistical properties that contribute to a strong initialization for downstream tasks.

As a consequence, while the pre-training loss can be used as an overall guide, due to the fact that certain hyper-parameters and architectural changes have a direct effect on the difficulty of the pre-training task, and subsequently the pre-training loss, the pre-training loss cannot be used as an indicator of downstream performance. To overcome this issue, recent studies have proposed using the rank of the output of SSL models for evaluation.

\subsection{Using rank for unsupervised evaluation}
\label{sec:related-work}
The rank of a matrix is the number of linearly independent rows or columns and is a proxy for the information content of the matrix. Thus, the higher the rank, the less redundancy, the more information is contained in the matrix. The classical way to compute the rank of a matrix is to count the number of non-zero singular values. However, when working with real-world matrices which contain non-integer values, many, if not all of the singular values will be non-zero. This necessitates the use of a threshold. However, in practice this threshold is quite application-dependent and difficult to set. To solve this issue, researchers proposed using the effective rank (E-Rank)~\cite{entropicRank}, which is a real-valued extension of the rank: 

\begin{equation}
\textrm{E-Rank}(A) = \exp \left( -\sum_{i=1}^{\min(N, M)}p_i\log p_i \right),
\label{eq:effective-rank}
\end{equation}
where $A$ is a matrix of size $(N,M)$, and $p_i$ are the normalized singular values. The effective rank, coined RankMe, was shown to be useful in the evaluation of self-supervised representations in computer vision~\cite{rankme}.

In the speech domain, researchers used the effective rank by adapting it to temporal data with RankMe-\emph{t}~\cite{aldeneh2024towards} which can be calculated as follows. Given a dataset of $n$ audio files, we use an SSL model to generate series of embedding sequences $S$. Each $\mathbf{e}_i \in S$ is an embedding sequence with a respective length $T_i$, i.e. $\mathbf{e}_i = [\mathbf{e}_{i}^1, \mathbf{e}_{i}^2, ..., \mathbf{e}_{i}^{T_i}]$. Each sequence is summed along the time dimension resulting in one vector per audio file, $\mathbf{e}_i^{sum} = \sum_{t=1}^{T_i}\mathbf{e}_{i}^t$. These vectors are then concatenated together to form the matrix $Z = [\mathbf{e}_1^{sum}, \mathbf{e}_2^{sum}, ..., \mathbf{e}_n^{sum}]$. RankMe-\emph{t} is then calculated by taking the E-Rank of $Z$ or $\textrm{RankMe-\emph{t}}(Z) = \textrm{E-Rank}(Z)$.

Although RankMe-\emph{t} was shown to correlate well with downstream performance~\cite{aldeneh2024towards}, correlations with the pre-training loss values were not reported, leaving unanswered whether RankMe and RankMe-\emph{t} serve as better indicators. Moreover, RankMe-\emph{t} operates at the utterance level as each sequence of embeddings is summed together, resulting in one vector per utterance. While this might capture useful information as to whether or not the model is pre-training well, we hypothesis that RankMe-\emph{t} looses useful information.

We build on and expand prior research by experimenting with cluster measures, investigating the use of rank and cluster measures in the early stages of pre-training, and focusing on two different tasks which are ASR and SV.

\section{Methods}
\label{sec:experiments}
In this section, we introduce the unsupervised measures used in our experiments and then describe the experimental settings, including the models trained and the downstream tasks.

\subsection{Unsupervised measures for SSL model evaluation}

We hypothesize that the dispersion (or lack thereof) of the output embeddings in the latent space is correlated with the downstream performance of the model. In order to test this hypothesis, we use the following clustering and rank-based measures.

\subsubsection{Clustering measures}

For clustering measures, we train KMeans~\cite{kmeans2, kmeans} clustering models on the SSL embeddings from a given layer. Specifically, we use MiniBatchKmeans~\cite{minibatchkmeans}, a faster variant of KMeans implemented in the scikit-learn~\cite{scikit-learn} library, with a KMeans++ initialization~\cite{kmeans++}. We set $k=1024$ as the number of clusters. 


Once a clustering model is trained, we use two metrics for evaluating the clustering quality: inertia, also called the Within-Cluster Sum of Squares~(WCSS), and the Davies-Bouldin~(DB) index~\cite{db_score}. The WCSS measures the average intra-cluster distance, or the mean distance between each data point and its assigned cluster centroid. This can be expressed by the following:

\begin{equation}
\textrm{WCSS} = \sum_{k=1}^{K} \sum_{i \in C_k} || x_i - \mu_k || ^2,
\end{equation}
where $K$ is the number of custers, $C_k$ is the set of data points assigned to cluster $k$, $x_i$ is a single data point and $\mu_k$ is the centroid of cluster $k$. 
In contrast to the WCSS, the DB index takes into the intra- and inter-cluster distance and is calulated by the following: 
\begin{equation}
\textrm{DB} = \frac{1}{K}\sum_{i=1}^{K}\max_{i \neq j}\frac{s_i + s_j}{d_{ij}},
\label{eq:db-index}
\end{equation}
where $K$ is the number of clusters, $s_i$ is the diameter of cluster $i$ and $d_{ij}$ is the distance between centroids of clusters $i$ and $j$. 


\subsubsection{Rank-based metrics}

For our rank based methods, we use RankMe-\emph{t} as described in Section~\ref{sec:related-work} as a baseline, and propose an additional rank based measure that we call
the Global Effective Rank~(GER).

Given the series of $n$ embedding sequences $S$ (same as Section~\ref{sec:related-work}), let $C$ represent the concatenation of all the embeddings in $S$. We propose the Global Effective Rank as the Effective rank of $C$. This can be expressed as follows: 

\begin{equation}
C = \big\Vert_{i=0}^{n} S_i = \mathbf{e}_0 \Vert \mathbf{e}_1 \Vert \mathbf{e}_2 \Vert \cdots \Vert \mathbf{e}_n.
\label{eq:concat}
\end{equation}

\begin{equation}
\textrm{GER} = \textrm{E-Rank}(C).
\label{eq:global-effective-rank}
\end{equation}
The motivation behind GER is to attempt to capture better the true rank of the raw representations instead of computing the rank of a compressed version as is done in RankMe-\emph{t}.

\subsection{Experimental settings}
As previously noted, for our experiments we work with BEST-RQ due to its state-of-the-art performance, efficiency, and open source implementation. The efficiency is particularly important as we need to train many models to evaluate the methodology. We create various hyper-parameters configurations by varying the batch size from 15 to 120 minutes and the proportion of audio masked from 40\% to 80\%. We train all models for 200k steps on the 960 hours set of LibriSpeech~\cite{panayotov2015librispeech}. We focus on two main checkpoints: 50k steps and 200k steps. While most SSL speech models are trained for longer~\cite{baevski2020wav2vec, hsu2021hubert, pmlr-v162-chiu22a}, previous work has shown 200k steps to be sufficient for evaluation~\cite{parcollet2023efficiency, whetten2024open}.


For our first task, we evaluate the frozen SSL models on in-domain ASR using the ContextNet LibriSpeech-100 task from the MP3S benchmark~\cite{zaiem23b_interspeech}. In this task, a minimal ContextNet~\cite{han20_interspeech} probe is trained on the \emph{train-clean-100} subset of LibriSpeech, and performance is measured by word error rate (WER) on the \emph{dev-clean} set using greedy decoding without a language model.
Since fine-tuning is common in practice, we also fine-tune the SSL models on in-domain ASR data using the SpeechBrain fine-tuning recipe. Evaluation follows the same setup as the frozen model, measuring WER on the \emph{dev-clean} set with greedy decoding and no language model.
For our third task, we evaluate the frozen SSL models on out-of-domain ASR using the LSTM CommonVoice 11.0 Basque ASR task from MP3S. For this task, a 2-layer BiLSTM probe is trained on the provided train splits. Evaluation is done using the WER on the provided development set.
Lastly, to assess performance on a non-ASR task, we follow the MP3S benchmark to evaluate the frozen SSL models on Speaker Verification using VoxCeleb1~\cite{nagrani17_interspeech}. In this task, a ECAPA-TDNN~\cite{desplanques20_interspeech} probe is trained and models are evaluated using error rate on the development set.

\begin{table*}[!ht]
\caption{Pearson Correlations between downstream performance and various unsupervised measures. Compared to pre-training loss (first row), unsupervised metrics tend to exhibit a much stronger correlation with downstream performance at either 50k or 200k steps of pre-trainin. Highest correlations are found in the in-domain tasks. Measures based on rank are more correlated with downstream performance than clustering measures. LS = LibriSpeech, CV = CommonVoice, VoxCel = VoxCeleb, (f.t.) = SSL Model is fine-tuned.}
    \centering
    \label{tab:corr}
        \begin{tabular}{l@{\hskip 3pt}cccccccccccc}
        \toprule
        \textbf{Measure} & \textbf{LS 50k} & \textbf{LS 200k} & \textbf{LS 50k} (f.t.) & \textbf{LS 200k} (f.t.) & \textbf{CV 50k} & \textbf{CV 200k} & \textbf{VoxCel 50k} & \textbf{VoxCel 200k} \\
        
        \midrule
        Pre-train Loss & -0.183 & -0.303 & -0.170 & -0.332 & 0.107 & 0.143 & 0.084 & -0.598 \\
        \midrule
        Inertia (WCSS) & 0.686 & 0.622 & 0.710 & 0.633 & 0.412 & -0.556 & -0.000 & 0.367 \\
        Davies-Bouldin & -0.903 & -0.494 & -0.929 & -0.486 & -0.854 & -0.414 & 0.155 & 0.356 \\
        \midrule
        RankMe-t & -0.752 & -0.735 & -0.734 & -0.756 & -0.387 & 0.110 & 0.376 & 0.567 \\
        Global Rank (GER) & -0.931 & -0.630 & -0.962 & -0.634 & -0.885 & -0.003 & 0.163 & 0.584 \\
        \midrule
        In-Domain & Yes & Yes & Yes & Yes & No & No & No & No \\
        Epochs on DS task & 5 & 20 & 5 & 20 & 5 & 20 & 20 & 20 \\
        \bottomrule
        \end{tabular}
        
\end{table*}

\begin{table}
    \caption{Pearson Correlations between various unsupervised measures and \textbf{future} downstream performance. The unsupervised measures are calculated from the embeddings of SSL models at 50k pre-training steps and the correlation is measured with their respective final performances at 200k pre-training steps. Correlations for in-domain ASR remains high, but not for out-of-domain tasks. LS = LibriSpeech, CV = CommonVoice, VoxCel = VoxCeleb, (f.t.)~=~SSL Model is fine-tuned.}
    \centering
    \label{tab:pred-correlations}
    \begin{tabular}{l@{\hskip 3pt}cccc}
    \toprule
     \textbf{Measure} & \textbf{LS} & \textbf{LS (f.t.)} & \textbf{CV} & \textbf{VoxCel} \\
    \midrule
    Pre-train Loss & -0.296 & -0.324 & 0.140 & -0.604 \\
    \midrule
    Inertia (WCSS) & 0.692 & 0.729 & -0.378 & -0.027 \\
    Davies-Bouldin & -0.840 & -0.837 & 0.095 & 0.056 \\
    \midrule
    RankMe-t & -0.771 & -0.785 & 0.201 & 0.200 \\
    Global Rank (GER) & -0.903 & -0.905 & 0.159 & 0.318 \\
    \bottomrule
    \end{tabular}
\end{table}


\begin{figure}[t]
    \centering
    \begin{subfigure}[t]{0.22\textwidth}
        \centering
        \includegraphics[width=\textwidth]{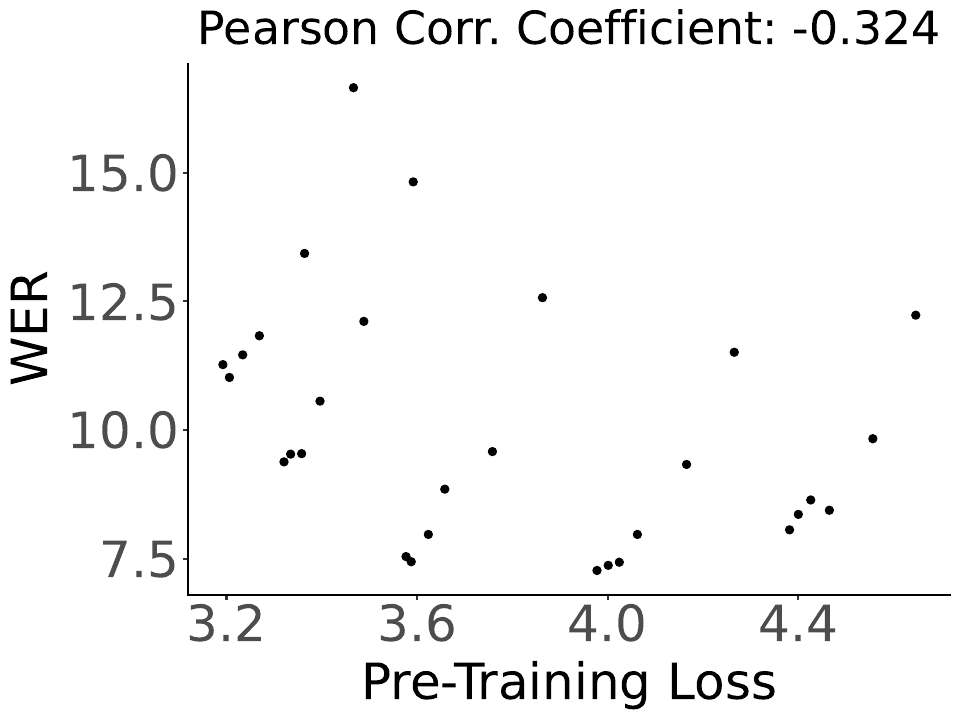}
        \caption{WER vs Pre-Training Loss}
        \label{fig:sub1}
    \end{subfigure}
    \begin{subfigure}[t]{0.22\textwidth}
        \centering
        \includegraphics[width=\textwidth]{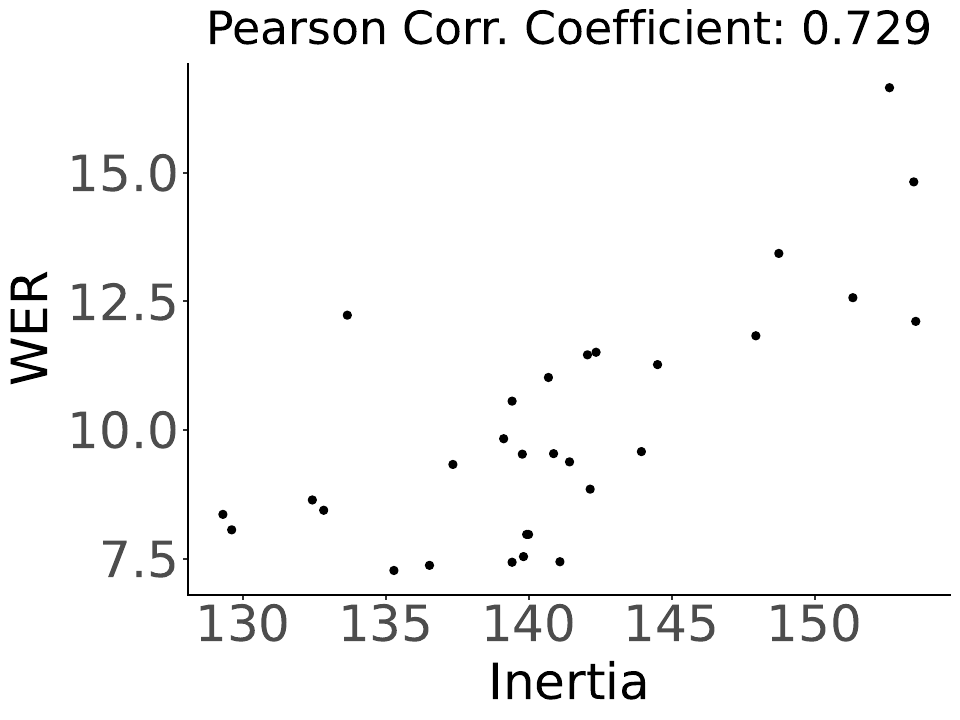}
        \caption{WER vs Inertia}
        \label{fig:sub2}
    \end{subfigure}
    \vspace{0.25cm} 
    \begin{subfigure}[t]{0.22\textwidth}
        \centering
        \includegraphics[width=\textwidth]{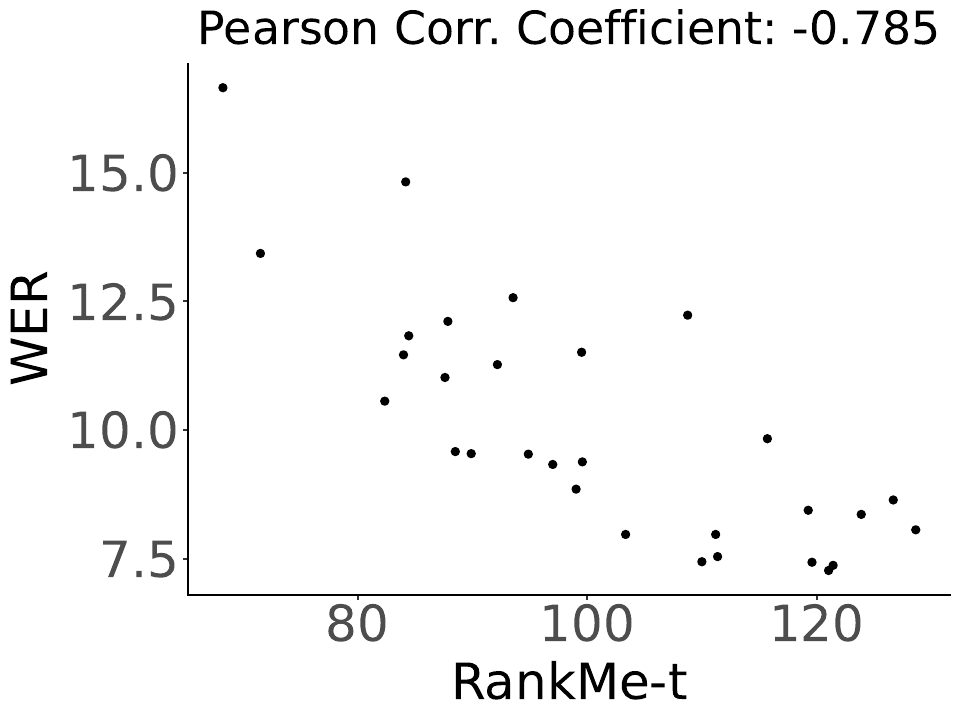}
        \caption{WER vs RankMe-t}
        \label{fig:sub3}
    \end{subfigure}
    \begin{subfigure}[t]{0.22\textwidth}
        \centering
        \includegraphics[width=\textwidth]{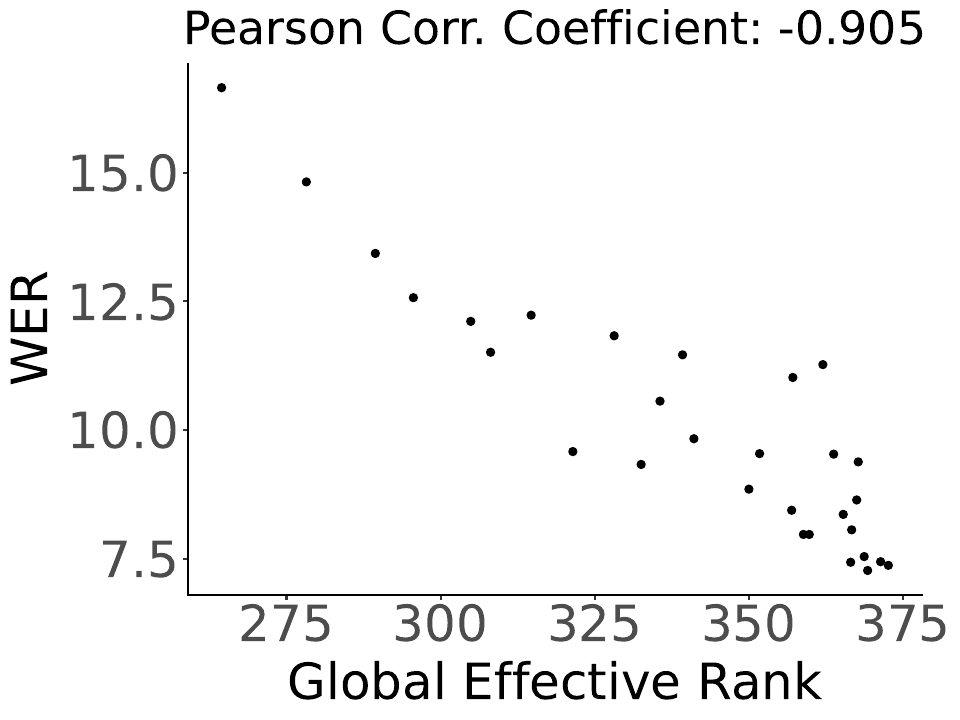}
        \caption{WER vs GER}
        \label{fig:sub4}
    \end{subfigure}
    \caption{
    Correlations between 4 selected measures (calculated on embeddings at 50k pre-training steps) and the future WER on the LibriSpeech \emph{dev-clean} set (at 200k pre-training steps and 20 epochs fine-tuning).
    Each point represents a model with different hyper-parameters (batch size and mask percentage). Fig. \ref{fig:sub1}, show that some of the models with the lowest WER (y-axis) have a relatively high pre-training loss (x-axis) resulting in a low correlation, whereas unsupervised measures have a stronger correlation with WER.
    }
    \label{fig:wer-vs-measures}
\end{figure}

\section{Results}
\label{sec:results}

We first report on preliminary experiments that informed our choices on the amount of audio to use for the unsupervised evaluation methods and the number of epochs for the downstream task. Then we present results for each task.

\subsection{Preliminary experiments}
For generating the output embeddings, we randomly sample about 1 hour of audio from each dataset. This stems from preliminary experiments in which we find that the global effective rank plateaus quickly, increasing minimally when using more than 1 hour of audio.  
Computing all embeddings and metrics for a single layer of one model takes about 20 minutes on CPUs, with the clustering accounting for most of the time.

Another preliminary finding is a strong correlation between the WER early on in pre-training and the final WER when working with in-domain data. For example, the Pearson correlations between WER from minimally trained models (50k pre-training steps and 5 downstream epochs) and their final counterparts (200k pre-training steps, 20 downstream epochs) were 0.951 and 0.967 when freezing and fine-tuning the SSL model, respectively. This means that in order to gauge performance with labeled data, one does not need to fully pre-train nor fully fine-tune when working with in-domain ASR. Thus, for the ASR experiments we only run the downstream for 5 epochs for the models at 50k checkpoints to save computation.

\subsection{Automatic speech recognition}
For in-domain ASR experiments, results were similar whether the SSL models were fine-tuned or frozen; both cluster and rank methods correlate with the WER better than the pre-training loss baseline (see first four columns of Table~\ref{tab:corr}). Some of the highest correlations came from the GER ranging from -0.63 to -0.96, meaning the higher the rank the better the performance (lower WER). In contrast, the pre-training did not correlated strongly with WER, with correlations ranging from -0.170 to -0.303. Graphs comparing a select few of these measures with the WER are shown in Fig.~\ref{fig:wer-vs-measures}.


Similarly, we find high correlations between clustering and rank measures at 50k steps of pre-training and final WER after 200k steps of pre-training and 20 epochs of fine-tuning. This sugests that, without any fine-tuning, we can guage the finial performance very early on in pre-training for in-domain ASR. For example, the correlations between the GER and final WER were around -0.9, yet only around -0.3 for the pre-training loss (first two columns of Table~\ref{tab:pred-correlations}).

When it comes to out-of-domain (CommonVoice) ASR experiments, correlations between unsupervised measure and WER remain high at 50k steps, with the highest being -0.885 for the GER. However, the correlations at 200k are not strong, nor are the correlations between unsupervised measures at 50k and the final WER at 200k (Table~\ref{tab:pred-correlations}). Examining why this might be the case, we observe that at 200k all the models perform similarly, with an absolute difference in WER of only 2.8 between the worst and best performance, suggesting that if model performance is very similar, it becomes hard to judge based on clustering or rank measures.

\subsection{Speaker verification}
With regards to Speaker Verification, we observe correlations between unsupervised measures and the error rate after 20 epochs of fine-tuning. While they are not as strong as with ASR, the unsupervised measures still correlate better than the pre-training loss. Note that for 200k checkpoints, there is a strong negative correlation between the loss and the downstream performance of -0.598 (last column Table~\ref{tab:corr}), indicating that the higher the loss, the lower the error rate which is not useful. We find that, like CommonVoice, the majority of models performed similarly (90\% of the models exhibited error rates within a 1\% range), leading to low correlations. 


\subsection{Additional observations} 
Although previous work has demonstrated that the most important layers for ASR are the middles layers as they have been shown to contain the most phonetic and word information~\cite{9688093}, our work agrees with~\cite{aldeneh2024towards} where the layers most useful for the downstream task are not the layers that correlated most with the downstream performance. For ASR, we found that the measures computed on the embeddings extracted from the first and last layers correlated the most with downstream performance, indicating that the way the model processes incoming acoustic features (or reconstructs them in the case of the last layers) is an important indicator of downstream ASR performance. For SV, we found that the middle layers tended to have stronger correlations than beginning or final layers. We report the correlations from measures calculated on the outputs of layer 12 for ASR and layer 8 for SV in both Table~\ref{tab:corr} and~\ref{tab:pred-correlations}. 

One other finding is that the correlations between inertia and downstream performance are positive except in the case of out-of-domain ASR at 200k steps of pre-training where all the models converged to very similar performance. This indicates that the lower the inertia, the better the clustering, and thus better performance. On the other hand, the correlations with the DB index are negative for ASR indicating the \emph{worse} the clustering, the better the performance. We believe that this because the DB index penalizes clusters that are close to their nearest neighbors. However, overlapping clusters do not necessarily indicate poor performance; instead, close neighboring clusters suggest efficient use of the latent space.

\section{Discussion}
\label{sec:discussion}
We propose using clustering and rank measures as efficient, unsupervised methods for evaluating the performance of SSL speech models and show that these measures are better indicators of downstream performance than the pre-training loss. To illustrate the potential compute
savings, suppose one pre-trains the 30 models from this work to 400k steps. With our settings this would take around 16,800 GPU hours. If one instead pre-trained all models to 50k and then used the rank to select and train only the top 5 models to 400k steps, this would take approximately 6,000 hours, saving approximately 10,800 GPU hours. 
We believe that future work could involve exploring these methods further and developing more robust measures to accurately evaluate SSL speech models early on in training.
We believe that future work could involve exploring these methods further and developing.

\section{Acknowledgements}
This work received funding from the French ANR E-SSL project N°ANR-22-CE23-0013 and used HPC resources from GENCI–IDRIS projects AD011014732 and A0131013821.

\bibliographystyle{IEEEtran}
\bibliography{mybib}

\end{document}